\documentclass[amsmath,amssymb,
 aps,
twocolumn
]{revtex4}

\newcommand{\lsim}   {\mathrel{\mathop{\kern 0pt \rlap
  {\raise.2ex\hbox{$<$}}}
  \lower.9ex\hbox{\kern-.190em $\sim$}}}
\newcommand{\gsim}   {\mathrel{\mathop{\kern 0pt \rlap
  {\raise.2ex\hbox{$>$}}}
  \lower.9ex\hbox{\kern-.190em $\sim$}}}

\usepackage{graphicx}
\usepackage{dcolumn}
\usepackage{bm}
\usepackage[utf8]{inputenc}
\usepackage{slashed}
\usepackage{physics}
\usepackage[colorlinks=true,linktoc=all]{hyperref}
\usepackage{xcolor}
\usepackage{changebar}
\usepackage{mathtools}
\newcommand{\Od}{{\cal O}}

\DeclareMathOperator{\cn}{cn}

\begin{document}

\title{Vector dark radiation and gravitational-wave polarization}

\author{Alfredo D. Miravet}
 \email{alfrdelg@ucm.es}
\affiliation{Departamento de F\'{\i}sica Te\'orica and Instituto de F\'isica de Part\'iculas y del Cosmos (IPARCOS), Universidad Complutense de Madrid, 28040 Madrid, Spain}
\author{Antonio L. Maroto}
\email{maroto@ucm.es}
\affiliation{Departamento de F\'{\i}sica Te\'orica and Instituto de F\'isica de Part\'iculas y del Cosmos (IPARCOS), Universidad Complutense de Madrid, 28040 Madrid, Spain}


\date{\today}

\begin{abstract}
We consider conformal vector models which could play the role of a cosmological dark radiation component. We analyse the propagation of gravitational waves in the presence
of this vector background and find a suppression in the tensor transfer function at large scales. We also find that although the cosmological background metric is isotropic, anisotropies are imprinted in the tensor power spectrum. In addition, the presence of the background vector fields induces a net polarization of the gravitational wave background and, for certain configurations of the vector field, a linear to circular polarization conversion. We also show that this kind of effects are also present for vector models with more general potential terms.

\end{abstract}

\maketitle


\section{Introduction}\label{sec:Intro}

Models containing additional vector fields have been proposed  in different cosmological contexts in recent years. From the pioneering works on inflation 
driven by vector fields \cite{Ford}  to the more recent ones based on non-abelian gauge fields (see \cite{Maleknejad:2012fw} and references therein), the vector inflation models have shown a rich phenomenology including the possible generation of primordial vector modes or statistical anisotropies in the CMB power spectrum \cite{Watanabe:2010fh,Ackerman:2007nb}. Vector fields have also been proposed as candidates for dark energy, either from potential terms \cite{Armendariz-Picon:2004say,Boehmer:2007qa} or from purely kinetic actions \cite{BeltranJimenez:2008iye,BeltranJimenez:2008enx}. More recently the possibility of constructing models of ultralight dark matter from 
coherently oscillating massive vector fields have been also analysed in detail both at 
the background \cite{Nelson:2011sf, LopezNacir:2018epg} and perturbation levels \cite{Cembranos:2016ugq}. This kind of models
exhibit a completely new phenomenology compared to the more standard ultralight  dark matter models based on scalar fields. Thus, apart from the suppression of the matter power spectrum on small scales which is typical of any fuzzy dark matter model, the presence of the background vector field induces a mixing between scalar, vector and tensor modes which
allows the generation of gravitational waves (GWs) from the usual density perturbations \cite{Cembranos:2016ugq}. In addition, the propagation of tensor modes is also  modified with respect to standard General Relativity inducing an anisotropic suppression of the tensor power spectrum on large scales \cite{Miravet:2020kuj}.

However, apart from dark matter or dark energy, it is also possible to employ vector fields to model other types of perfect 
fluids. Thus, a general result proven in \cite{Cembranos:2012kk,Cembranos:2012ng,Cembranos:2013cba} shows that coherently oscillating homogeneous fields of arbitrary spin will have an 
isotropic average energy-momentum tensor provided that the oscillations are faster than the universe expansion rate. The average equation of state will depend on the type of potential term driving the oscillations.  Thus in particular, a mass term (quadratic potential) leads to a pressureless fluid as mentioned before, but in general a $V(A)=\lambda (A_\mu A^\mu)^n$ type potential will lead to a 
$w=(n-1)/(n+1)$ equation of state. Precisely in this work we will explore the possibility of constructing vector models of dark radiation, typically with quartic potentials,  and explore some of their phenomenological implications. 

Dark radiation is the possible radiation component existing in the universe in addition to the standard 
radiation content corresponding to photons and
neutrinos. The abundance of dark radiation is usually parametrized through the effective number of neutrino species $N_\text{eff}$
defined as 
\begin{align}\label{eq:rhoRNeff}
\rho_R=\rho_\gamma+N_{\text{eff}}\frac{\pi^2}{30}\frac{7}{8}\left(\frac{4}{11}\right)^{4/3}T^4,
\end{align}
where $\rho_R$ is the total radiation density and $\rho_\gamma$ is the 
photon energy density. 
The Standard Model prediction is $N^{SM}_{\text{eff}}=3.046$ so 
that the abundance of dark radiation is parametrized by
\begin{align}\label{eq:DeltaNeff}
\Delta N_{\text{eff}}=N_{\text{eff}}-3.046.  
\end{align}
The current limits on this parameter from Planck 2018 TT+TE+EE+lensing and BAO reads \cite{Planck:2018vyg}
\begin{align}\label{eq:Neffbound}
\Delta N_{\text{eff}}<0.28\; (95\% \text{C.L.}).  
\end{align}

Future Stage-4 CMB experiments will increase the sensitivity up to $\Delta N_{\text{eff}}\simeq 0.03$ \cite{CMB-S4:2016ple}. Dark radiation increases the expansion rate of the universe in the radiation era, thus reducing the size of the sound horizon at
recombination. This can be compensated by an increase in the Hubble parameter today $H_0$ so that the CMB temperature power spectrum remains unchanged \cite{Buen-Abad:2019opj}. 
Thus, dark radiation has been proposed as a possible way to alleviate the 
$H_0$ tension between local universe and CMB observations \cite{Riess:2019cxk}. The effect of dark radiation on inflation-produced gravity waves has been studied in \cite{Jinno:2012xb}.

Dark radiation is usually described by means of new relativistic particles weakly interacting with the Standard Model sector. Thus, models based on axion-like particles have been proposed in \cite{Conlon:2013isa, Marsh:2015xka}, supersymmetric candidates have also been considered related to the cosmological gravitino problem or axino decays \cite{Hasenkamp:2011em, Ichikawa:2007jv} among others.

In order to construct  dark radiation models from cosmological vector fields, we will focus on conformal vector models either for abelian or non-abelian fields \cite{Gomez:2021jbo}. Such models \cite{Asorey:2021rwv} are generically described  by Maxwell or Yang-Mills terms plus a quartic self-interaction potential. Even though at the background level the vector field configurations we will consider  behave as isotropic perfect fluids, the presence of the background vector fields induces non-vanishing anisotropic stresses in the perturbed energy-momentum tensor. As shown in \cite{Miravet:2020kuj}, such anisotropic stresses modify the propagation equation of gravitational waves. Notice  that this type of effect is not generated by cosmological scalar fields. 
An important consequence of this modification that we will analyse in this work is that, unlike the case of massive vector fields \cite{Miravet:2020kuj}, vector dark radiation can induce a net polarization
of the primordial gravitational wave background. This would be a clear smoking gun of this kind of models. Indeed, the primordial background of gravitational waves generated during inflation is expected to be unpolarized in standard cosmology. Nonetheless, it is possible to generate circularly polarized primordial gravitational waves in extended versions of inflation, for  example with Chern-Simons gauge or gravitational couplings of the inflaton \cite{Lue:1998mq, Sorbo:2011rz}. Primordial helical turbulence produced in first-order phase transition has been also proposed as a mechanism for  the generation of gravitational wave circular polarization \cite{Kahniashvili:2005qi}. After inflation, polarization of the gravitational wave backgrounds, both astrophysical and cosmological,  can be induced by interaction with matter structures \cite{Cusin:2018avf},  
though the amount of polarization produced by this mechanism is relatively small.
Prospects for the detection of gravitational wave polarization with current and future detectors have been explored in \cite{Kato:2015bye, Domcke:2019zls, Sato-Polito:2021efu}.

The paper is organised as follows. Firstly, in Section \ref{sec:vectordr} we present our vector dark radiation model and study its dynamics in an expanding universe. In Section \ref{sec:GWs} we review the basics of GW propagation and GW backgrounds, and introduce the formalism of Stokes parameters to characterize their degree of polarization. In Sections \ref{sec:GWwithDR} and \ref{sec:polarized_bg} we explore the effect of our dark radiation model in the GW propagation and the impact on unpolarized and polarized GW backgrounds, respectively. In Section \ref{sec:general_potential} we obtain the GW propagation equations in presence of a vector field with a general potential. Finally, in Section \ref{sec:conclusions} we draw the main conclusions of our work.

\section{Vector Dark Radiation}\label{sec:vectordr}

In this section, we consider a simple model for dark radiation based upon a vector field $A_\mu$ with a quartic potential. The starting point is the action
\begin{equation}\label{eq:A4S}
    S = \int \dd^4 x \sqrt{-g} \left( -\frac{1}{4} F_{\mu\nu} F^{\mu\nu} - \frac{\lambda}{4}(A^2)^2  \right),
\end{equation}
where $A^2 = A_\mu A^\mu$, $g=\det(g_{\mu\nu})$ is the determinant of the  metric tensor, $F_{\mu\nu}=\partial_\mu A_\nu - \partial_\nu A_\mu$ is the field strength and $\lambda$ is a dimensionless parameter that determines the strength of the self-interaction.

The equations of motion for the vector field read
\begin{equation}\label{eq:A4eom}
    F^{\mu\nu}{}_{;\nu} + \lambda A^\mu A^2 = 0,
\end{equation}
where the semicolon denotes the covariant derivative. 
We will consider a flat Friedmann-Lemaître-Robertson-Walker (FLRW) background  metric  
\begin{align}
    \dd{s}^2 = a^2(\eta) [\dd{\eta}^2 - \delta_{ij}  \dd{x}^i \dd{x}^j],
\end{align}
and a homogeneous vector field $A_\mu(\eta)$, which depends solely on conformal time $\eta$. In these equations, Latin indices $i,j=1,2,3$ run over spatial components and, in the case of the metric perturbation $h_{ij}$, indices are raised and lowered with $\delta_{ij}$. We also use $\hbar=c=k_B=1$ units. It should be noted that, even though the background vector field breaks isotropy,
its average energy-momentum tensor will be isotropic as we will show below. Setting $\mu=0$ in \eqref{eq:A4eom}, we get 
\begin{equation}
    \lambda A_0 A^2 = 0.
\end{equation}

We are not interested in the lightlike solution $A^2=0$, since it would simply grow linearly in time, and as shown in \cite{Cembranos:2012kk} a fast oscillation around the potential minimum is necessary to ensure that anisotropic pressures average out. This necessarily sets $A_0 = 0$. Therefore, the equation with $\mu = i$ reads
\begin{equation}\label{eq:A4eomi}
    A_i'' + \lambda \vb{A}^2 A_i = 0,
\end{equation}
where $'\equiv\dd/\dd{\eta}$ and with $\vb{A}^2=\delta^{ij} A_i A_j$ the squared modulus of the spatial part of the vector field. This equation of motion allows for several configurations of the vector field, of which we shall analyse two particular cases: Linear and circular polarizations.

\subsection{Linear polarization}\label{subsec:vectorlin}

In the case of a linearly polarized vector field, it evolves along a fixed direction which can be identified with the $z$-axis after a convenient orientation of the axes:
\begin{equation}
    A_\mu(\eta) = (0, 0, 0, A_z(\eta)). 
\end{equation}

Working out the spatial equations of motion \eqref{eq:A4eomi} in components, we get
\begin{equation}
    A_z'' + \lambda A_z^3 = 0.
\end{equation}
which agrees with the corresponding equation in flat space-time thanks to the conformal
invariance of the action \eqref{eq:A4S}. 
An analytic solution in terms of Jacobi elliptic functions \cite{Finkel:2000nu} can be readily obtained. Assuming an initial value $A(\eta_\text{in})=A_\text{in}$ with zero derivative and setting $\eta_\text{in} = 0$ for simplicity, the solution is given by
\begin{equation}\label{eq:A4cn}
    A_z(\eta) = A_\text{in} \cn\left(\sqrt{\lambda}A_\text{in}\eta;1/2\right),
\end{equation}
where $\cn(x;m)$ is the elliptic cosine function with square modulus $m$. This function is periodic on its first argument, with period $4K(m)$, where $K(m)$ is the complete elliptic integral of the first kind, and in particular $K(1/2)\simeq 1.854$. Thus, the field has got a naturally associated comoving frequency, given by
\begin{equation}\label{eq:freq}
    \omega = \sqrt{\lambda}A_\text{in}.
\end{equation}

Such frequency can be compared with the expansion rate of the Universe, given by the comoving Hubble parameter $\mathcal{H}=a'/a$, so that if the condition $\omega\gg\mathcal{H}$ is satisfied, the average energy-momentum tensor becomes isotropic.

The stress-energy tensor  obtained from  the action \eqref{eq:A4S}  reads
\begin{equation}\label{eq:A4T}
    T^\mu{}_\nu = \frac{1}{4}\left[ F_{\rho\sigma} F^{\rho\sigma} + \lambda (A^2)^2\right] \delta^\mu{}_\nu - F^{\mu\rho} F_{\nu\rho} - \lambda A^\mu A_\nu A^2.
\end{equation}

The energy density can then be calculated by plugging the analytical solution \eqref{eq:A4cn} into the stress-energy tensor
\begin{equation}
    \rho_A = T^0{}_0 = \frac{1}{2a^4}\left( A_z'^2 + \frac{\lambda}{2} A_z^4\right) = \frac{\lambda A_\text{in}^4}{4a^4},
\end{equation}
which as expected for a conformal theory scales exactly as radiation, i.e, $\rho_A\propto a^{-4}$. It is immediate then to obtain today's abundance
\begin{equation}
    \Omega_A = \frac{\rho_{A,0}}{\rho_c} = \frac{2\pi G}{3 H_0^2}\lambda A_\text{in}^4.
\end{equation}

The model is completely characterized at the background level by two parameters $(\omega, \Omega_A)$, i.e. the oscillation frequency  and the dark radiation abundance.
The current observational constraints on such parameters come, on one hand, from the limits on the effective number of neutrino species discussed in the introduction. Thus,  
\begin{equation}
    \Omega_A \leq \frac{\Delta N_\text{eff}}{N_\text{eff}^\text{SM}+\frac{16}{7}\left(\frac{11}{4}\right)^{4/3}}\Omega_R^\text{SM} < 0.024\Omega_R \simeq 2\cdot 10^{-6},
\end{equation}
where we have used the current limits on $\Delta N_\text{eff}$ given by Eq. \eqref{eq:Neffbound}.

On the other hand, constraints on the frequency $\omega$ come from the 
requirement of isotropy. Since the vector  points in the direction of the $z$ axis, the 
 pressures $p_i = -T^i{}_i$ can be different
\begin{subequations}
\begin{equation}\label{eq:pxpy}
    p_x = p_y = \frac{\rho_A}{3}\left[3 - 6 \cn^4(\omega\eta;1/2)\right],
\end{equation}
\begin{equation}\label{eq:pz}
    p_z = \frac{\rho_A}{3}\left[ 12 \cn^4(\omega\eta; 1/2) - 3\right].
\end{equation}
\end{subequations}

The pressures are oscillating around the isotropic configuration of
$p_x=p_y=p_z=\rho_A/3$ with a larger amplitude in the $z$ direction. Notice that the 
average of the term involving the elliptic function is $\langle \cn^4 \rangle = 1/3$.
In the regime of fast oscillations $\omega\gg {\cal H}$, it has been shown \cite{Cembranos:2016ugq} that the
 effect of the pressure oscillations on the background metric is suppressed by 
 ${\cal H}/\omega$. Therefore, for sufficiently large frequencies, the energy-momentum tensor can be replaced by the average isotropic tensor. In this scenario, the field can be described as a perfect fluid and the description of the spacetime with a FLRW metric is correct.

Since $\mathcal{H}$ is monotonically decreasing in radiation and matter-dominated epochs, once the field has entered the fast-oscillation regime, it will not leave it throughout its whole evolution afterwards. In particular, if it oscillates quickly at the end of reheating, when the radiation temperature is $T_\text{RH}$, i.e. provided  
\begin{equation}
    \omega\gg\mathcal{H}(T_\text{RH}) = 265\text{ Hz}\left(\frac{T_\text{RH}}{10^{10}\text{ GeV}}\right).
\end{equation}
then the field will be in the fast oscillation regime at all times afterwards. 

Even if the frequency is below this value, the field would meet the fast oscillation
regime at a later time. In order to ensure a standard isotropic evolution from the time 
of nucleosynthesis and recover the observed abundances of light elements in the Universe, 
 the corresponding condition reads
\begin{equation}
    \omega\gg\mathcal{H}(T_\text{nuc}) = 1.4\cdot 10^{-11}\text{ Hz}\left(\frac{T_\text{nuc}}{\text{MeV}}\right),
\end{equation}
with the nucleosynthesis temperature being around $T_\text{nuc}\sim 0.1$ MeV. In any case, if the anisotropies generated by the vector field, which can be roughly estimated as $\Omega_A/\Omega_R$, are smaller than the typical amplitude of anisotropies of the CMB, i.e. $\Od(10^{-5})$  one should not worry about the fast-oscillation condition.

\subsection{Circular polarization}\label{subsec:vectorcirc}

The linearly polarized solution constrains the oscillation of the field to a single direction, but that does not need to be the case. Another simple solution can be obtained by fixing the modulus of the field to be comovingly constant, i.e. $\vb{A}^2=\alpha^2$, with $\alpha$ a real constant. Under this condition, the equations of motion read
\begin{equation}
    A_i''+\lambda\alpha^2 A_i=0,
\end{equation}
which has a solution in terms of trigonometric functions, so that the vector field revolves in a circular motion, with frequency
\begin{equation}
    \omega=\sqrt{\lambda}\alpha.
\end{equation}

If we choose the $z$-direction to be perpendicular to the rotation plane, and the vector field to initially point towards the $x$-direction, the particular solution can be written as
\begin{equation}\label{eq:A_circ}
    \vb{A}(\eta) = \alpha(\cos\omega\eta,\sin\omega\eta,0).
\end{equation}
Notice that this solution is compatible with 
the initial ansatz $\vb{A}^2=\alpha^2$.

The stress-energy tensor is still given by \eqref{eq:A4T}, though both energy density and pressures are different due to the different solution. On the one hand, the energy density is homogeneous and given by
\begin{equation}
    \rho_A = \frac{3\lambda\alpha^4}{4a^4},
\end{equation}
so that today's abundance is
\begin{equation}
    \Omega_A = \frac{2\pi G \lambda \alpha^4}{H_0^2},
\end{equation}
which together with the frequency $\omega$ can be used as the two parameters that characterise our model. On the other hand, the pressures $p_i = -T^i{}_i$ are given by
\begin{subequations}
\begin{equation}\label{eq:pxpyc}
    p_{x} = \frac{\rho_A}{3}\left[1+ 4\cos(2\omega\eta)\right],
\end{equation}
\begin{equation}
    p_{y} = \frac{\rho_A}{3}\left[1- 4\cos(2\omega\eta)\right],
\end{equation}
\begin{equation}\label{eq:pzc}
    p_z = \frac{\rho_A}{3},
\end{equation}
\end{subequations}
and the non-vanishing anisotropic pressures are
\begin{equation}
    T^x{}_y = T^y{}_{x} = -\frac{\rho_A}{3}\sin(2\omega\eta).
\end{equation}

As in the linearly polarized case, the pressures oscillate around the homogeneous configuration of a radiation component, given by $p_i=\rho/3$ and vanishing anisotropic pressures. In the fast-oscillation regime $\omega\gg\mathcal{H}$ these deviations average out, so the discussion in the previous section regarding the value of the frequency $\omega$ and the isotropy of the Universe can also be applied here.

\section{Tensor Power Spectra and Stokes Parameters}\label{sec:GWs}

Before we move to the analysis of the effects of the background vectors on gravitational wave propagation, we will briefly review the fundamental quantities employed to describe
the amplitude and polarization of gravity waves. Let us thus consider a flat FLRW metric with tensor perturbations
\begin{equation}\label{eq:FLRWh}
    \dd{s}^2 = a^2(\eta) [\dd{\eta}^2 - (\delta_{ij} + h_{ij}) \dd{x}^i \dd{x}^j],
\end{equation}
where $h_{ij}$ is the transverse and traceless gauge-invariant tensor perturbation, satisfying
\begin{equation}
    h_{ii}=\delta^{ij}h_{ij}=0,\qquad \partial_i h_{ij}=0.
\end{equation}

After Fourier-transforming the tensor metric perturbation, if we work in a frame $\{\vb{u}_1,\vb{u}_2,\vb{u}_3\}$ so that $\vb{u}_3 = \vb{\hat k}$ is the propagation direction of the GW with wavevector $\vb{k}$, then this tensor can be written as
\begin{equation}\label{eq:hijmatrix}
	h_{ij}=\begin{pmatrix}
		h_+ & h_\times & 0\\
		h_\times & -h_+ & 0\\
		0 & 0 & 0
	\end{pmatrix},
\end{equation}
where $\{+,\cross\}$ refer to the linear polarization basis which can be easily related to the right-left circular polarization basis via
\begin{equation}\label{eq:hLR}
    h_{\substack{R \\ L}} = \frac{ h_{+} \mp i h_{\cross} }{\sqrt{2}}.
\end{equation}

The perturbed Einstein equation in the absence of GW sources $\delta G_{\mu\nu}=0$ for the metric \eqref{eq:FLRWh} yields the well-known GW propagation equation
\begin{equation}\label{eq:GWpropvacuum}
    h_\lambda''+2\mathcal{H}h_\lambda'+k^2 h_\lambda = 0,
\end{equation}
where $k = \abs{\vb{k}}$ and $\lambda = \{+,\cross\}$ is one of the polarizations. The evolution of a mode is entirely determined by its momentum $k$ and can be qualitatively described in the following way: If the mode is super-Hubble ($\mathcal{H}\gg k$), its amplitude remains constant, whereas when it enters the Hubble horizon ($\mathcal{H}\ll k$) it oscillates with its amplitude damped as $1/a$.

Given a stochastic background of GWs, as it is assumed to be the case for the GWs generated in the early universe, one can define the tensor power spectrum $P_T(\vb{k},\eta)$ in the usual way
\begin{equation}
    \sum_{\lambda,\lambda'}\langle h_\lambda(\eta,\vb{k}) h_{\lambda'}^* (\eta, \vb{k'}) \rangle = \delta^{(3)}(\vb{k}-\vb{k'}) P_T(\vb{k},\eta),
\end{equation}
where $\langle\dots\rangle$ represents an ensemble average. In a similar way, power spectra for both linear polarizations can also be defined, as well as a correlator between them
\begin{subequations}\label{eq:PplusPcross}
\begin{equation}
    \langle h_+(\eta,\vb{k}) h_{+}^* (\eta, \vb{k'}) \rangle = \delta^{(3)}(\vb{k}-\vb{k'}) P_+(\vb{k},\eta),
\end{equation}
\begin{equation}
    \langle h_{\cross}(\eta,\vb{k}) h_{\cross}^* (\eta, \vb{k'}) \rangle = \delta^{(3)}(\vb{k}-\vb{k'}) P_{\cross}(\vb{k},\eta),
\end{equation}
\begin{equation}
    \langle h_+(\eta,\vb{k}) h_{\cross}^* (\eta, \vb{k'}) \rangle = \delta^{(3)}(\vb{k}-\vb{k'}) P_{+\cross}(\vb{k},\eta).
\end{equation}
\end{subequations}

If linear polarizations are uncorrelated, i.e. $\langle h_+ h^*_{\cross}\rangle = \langle h_{\cross} h^*_+\rangle = 0$, then the total power spectrum is just the sum of the partial power spectra $P_T = P_+ + P_{\cross}$, whereas the difference $P_+ - P_{\cross}$ yields the net linear polarization of the GW background. If correlation does exist, its real part is associated with the linear polarization as well, whereas its imaginary part indicates parity violation.

In order to quantify these degrees of polarization, it is possible to define the Stokes parameters for the GWs \cite{Gubitosi:2016yoq,Seto:2008sr} $I$, $Q$, $U$ and $V$, which are analogous to their more common electromagnetic counterparts. These can be expressed in terms of the plus and cross power spectra and the plus-cross correlator \eqref{eq:PplusPcross} as
\begin{subequations}\label{eq:Stokes}
\begin{equation}
    I = P_+ + P_{\cross} = P_T,
\end{equation}
\begin{equation}
    Q = P_+ - P_{\cross},
\end{equation}
\begin{equation}
    U = -2\Re P_{+\cross},
\end{equation}
\begin{equation}
    V = -2\Im P_{+\cross}.
\end{equation}
\end{subequations}

$I$ is the total power spectrum, $Q$ accounts for linear polarization in the plus-cross basis, $U$ measures linear polarization in a basis that differs in a rotation from the former, and $V$ is the circular polarization. Non-zero $Q$ or $U$ parameters indicate an anisotropic GW configuration, whereas a non-zero $V$ parameter means a parity violating configuration.

The primordial GW background generated during inflation is typically described as a Gaussian, isotropic and unpolarized stochastic ensemble (even though each particular realisation does not need to be so individually), in which case the equality
\begin{equation}
    P_{+,\text{in}} = P_{\cross,\text{in}} = \frac{1}{2} P_{T,\text{in}}
\end{equation}
arises naturally, with ``in'' referring to primordial quantities. The primordial tensor power spectrum per logarithmic interval in $k$, denoted $\mathcal{P}_{T,\text{in}}$, is usually parametrised in terms of the tensor amplitude $A_T$ and tensor tilt $n_T$ as
\begin{equation}\label{eq:PTparametrization}
    \mathcal{P}_{T,\text{in}}(k) = \frac{k^3}{2\pi^2} P_{T,\text{in}}(k) = A_T(k_*) \left(\frac{k}{k_*}\right)^{n_T},
\end{equation}
where $k_*$ is the pivot scale. The power spectrum within standard cosmology can then be obtained at a later time  through the transfer function $T(k,\eta)$, which relates the value of a specific GW mode at a specific time with its primordial value
\begin{equation}\label{eq:Tdef}
    h_\lambda(k,\eta) = T(k, \eta) h_\lambda(k, \eta_\text{in}).
\end{equation}
Since power spectra relate via the squared modulus of the transfer function, one has
\begin{equation}\label{eq:PTSM}
    P_{T}(k, \eta) = |T(k,\eta)|^2 P_{T,\text{in}}(k).
\end{equation}
Notice that in standard cosmology, the transfer function 
does not depend on the propagation direction nor the polarization  of the GW,  
 as can be seen from the equation of propagation  \eqref{eq:GWpropvacuum}. As a result, if the GW background is primordially isotropic or unpolarized, these will be an ever-present feature of this background at any future time. 

\subsection{Modified propagation}

Even though the primordial background possesses these features, there is also the possibility that, in the framework of a theory beyond the standard cosmology, the modified  GW propagation equations differ for both polarizations, resulting in a richer power spectrum in later stages of cosmic evolution. The deviation from $\Lambda$CDM can then be encoded in four \textit{ratio functions} $R_{\lambda\lambda'}$, which are defined as the ratio of the GW amplitudes beyond (labelled \emph{new}) and within the standard model (labelled \emph{SM}) \cite{Miravet:2020kuj}
\begin{equation}
    \begin{pmatrix}
        h_+^\mathrm{new}\\
        h_\times^\mathrm{new}
    \end{pmatrix} = \begin{pmatrix}
        R_{+} & R_{+\times}\\
        R_{\times +} & R_{\times}
    \end{pmatrix}\begin{pmatrix}
        h_+^\mathrm{SM}\\
        h_\times^\mathrm{SM}
    \end{pmatrix},
\end{equation}
where we define $R_\lambda\equiv R_{\lambda\lambda}$. Note that non-zero $R_{\lambda\lambda'}$ with $\lambda\neq\lambda'$ are only possible when both polarizations mix, since one has to act as a source of the other. Besides cosmological parameters, these ratio functions can have additional dependencies such as the direction of observation or new parameters of the model.

Let us consider a theory for which both linear polarizations mix, a phenomenon that can affect all Stokes parameters. In the same way that the transfer function \eqref{eq:Tdef} relates the GW mode in two different moments, the ratio function can be thought of as a transfer function between two different cosmological models, the original one being standard cosmology. With this idea, we can write the Stokes parameters \eqref{eq:Stokes} for the new theory in terms of the standard tensor power spectrum \eqref{eq:PTSM} and the ratio functions as $S = \mathcal{S} P_T$, where $S$ is any of the four Stokes parameters and $\mathcal{S}$ the associated reduced Stokes parameter. These have the following expressions
\begin{subequations}\label{eq:reduced_stokes}
\begin{equation}
    \mathcal{I} = \frac{1}{2}(|R_{+}|^2 + |R_{\times}|^2 + |R_{+\times}|^2 + |R_{\times +}|^2),
\end{equation}
\begin{equation}
    \mathcal{Q} = \frac{1}{2}(|R_{+}|^2 - |R_{\times}|^2 + |R_{+\times}|^2 - |R_{\times +}|^2),
\end{equation}
\begin{equation}
    \mathcal{U} = -\Re(R_{+}R_{\times +}^* + R_{\times} R_{+\times}^*),
\end{equation}
\begin{equation}
    \mathcal{V} = -\Im(R_{+}R_{\times +}^* - R_{\times} R_{+\times}^*).
\end{equation}
\end{subequations}

It is clear from these equations that an  anisotropic configuration of the GW background can be achieved either by having a different propagation for each polarization $R_+\neq R_\times$ or a correlation between them $R_{\lambda\lambda'}\neq 0$. Parity violation requires this correlation to be complex.

\section{Effect of vector dark radiation on GW propagation}\label{sec:GWwithDR}

In order to analyse the effects of the background vector field, we need  to go to first order in metric perturbations and obtain the modified equation of propagation. 
The equation to solve is the transverse-traceless Einstein equation up to first order in metric perturbations for the metric \eqref{eq:FLRWh}, which reads
\begin{equation}
	\Lambda_{ij,lm}(\delta G^l{}_m-8\pi G\delta T^l{}_m)=0,
\end{equation}
with $\delta T^l{}_m$ being the perturbed energy-momentum tensor \eqref{eq:A4T},
\begin{equation}
	\Lambda_{ij,lm}=P_{il}P_{jm}-\frac{1}{2}P_{ij} P_{lm}
\end{equation}
is the TT projector and
\begin{equation}
	P_{ij}=\delta_{ij}-\hat{k}_i\hat{k}_j.
\end{equation}

Here we are only interested in the effects on propagation, so that we ignore perturbations of the vector field that would act as a source term of GWs, i.e. we only pay attention to those terms proportional to $h_{ij}$ in $\delta T^l{}_m$. From this point onwards, we have to look at the two configurations described in Section \ref{sec:GWs} separately, as they are going to introduce different terms in the GW propagation equation.

\subsection{Linearly polarized vector field}\label{subsec:GWlin}

Firstly, let us look at the linearly polarized vector field. Thanks to axial symmetry around the direction of observation, we choose a basis in which GWs travel along the $z$-axis and the vector field is contained in the $yz$-plane, so we arrive to the modified equations of propagation for GWs, which read
\begin{equation}
    h_{\lambda}''+2\mathcal{H}h_{\lambda}'+\left[k^2-\frac{8\pi G\sin^2\theta}{a^2}\left( \vb{A}'^2 - \lambda \vb{A}^4 L_\lambda \right) \right]h_{\lambda}=0,
\end{equation}
where $\vb{A}$ is the spatial part of $A_\mu$, $\theta$ is the angle between the direction of propagation and the vector field i.e. $\cos\theta = \hat{\vb{k}}\cdot\hat{\vb{A}}$ and
\begin{equation}
    L_\lambda = \begin{cases}
    1 + \sin^2\theta,&\lambda = +\\
    1,&\lambda = \times
    \end{cases}
\end{equation}
is a term that depends on the linear polarization mode. Here we see that the generation of a net GW linear polarization  is related to the anisotropy of the background vector field, since both equations differ as long as $\theta\neq 0$ even in the fast-oscillation regime of the field, which yields an isotropic stress-energy tensor at background level. Having different equations for both linear polarizations also implies a correlation between circular polarizations due to propagation, since their equations are no longer separable, but no net circular polarization is generated. Thus for the primordial stochastic background, we expect an anisotropic modification of intensity and linear polarization power spectra.

If we write the equations in terms of the analytic solution for the field obtained in Section \ref{subsec:vectorlin}, we get the following expressions
\begin{widetext}
\begin{subequations}\label{eq:hU1}
\begin{equation}
    h_{+}''+2\mathcal{H}h_{+}'+\left[k^2+\frac{6 H_0^2 \Omega_A \sin^2\theta}{a^2}\left( (3 + 2 \sin^2\theta) \cn^4(\omega\eta;1/2)-1  \right) \right]h_{+}=0,
\end{equation}
\begin{equation}
    h_{\cross}''+2\mathcal{H}h_{\cross}'+\left[k^2+\frac{6 H_0^2 \Omega_A \sin^2\theta}{a^2}\left( 3 \cn^4(\omega\eta;1/2)-1  \right) \right]h_{\cross}=0.
\end{equation}
\end{subequations}
\end{widetext}

\begin{figure*}[t]
    \centering
    \includegraphics{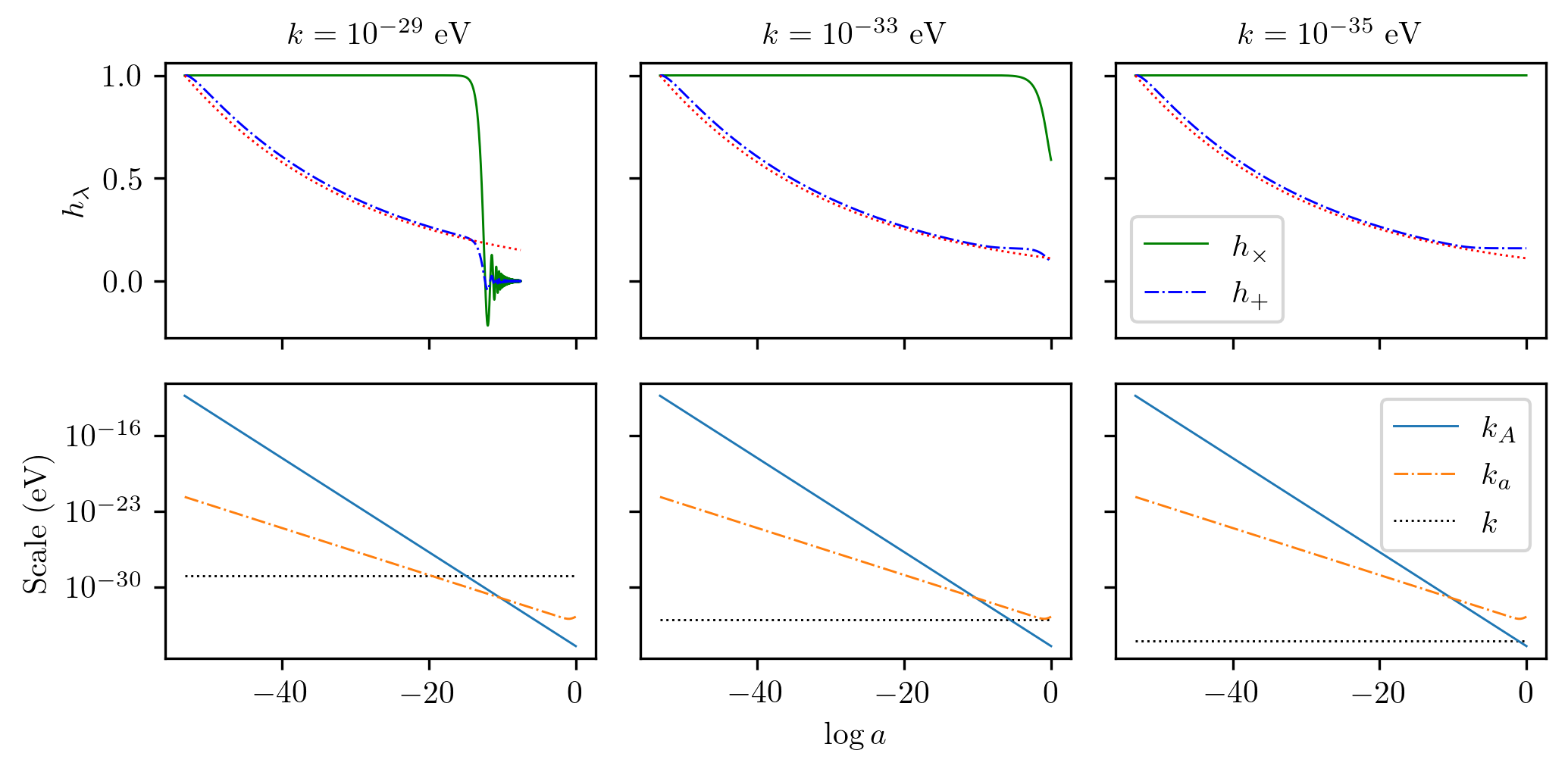}
    \caption{GW evolution for $\Omega_A = \Omega_R / 100$, $\omega=2500$ Hz, $\theta=\pi/2$ and three different wavenumbers, one in each column. The fast-oscillation regime applies throughout the entire evolution, so the results are valid for any frequency $\omega$ that satisfies so. The top row shows the evolution of the plus (dash-dotted) and cross (solid) modes, the later being just the standard evolution in $\Lambda$CDM, as well as the analytical solution for plus modes in a radiation-dominated epoch (dotted), given by Eq. \eqref{eq:hplusanalytical}. The modes are normalised by their primordial values. The bottom row shows the scales of the different terms in the propagation equation corresponding to the evolution above them, with $k_a=\sqrt{|a''/a|}$ the damping term scale and $k_A=\sqrt{6\Omega_A}H_0|\sin{\theta}|/a$ the vector field term scale.}
    \label{fig:evolution}
\end{figure*}

The amplitude of the new terms is proportional to the dark radiation abundance $\Omega_A$. In particular, one can expect a non-negligible effect so long as $6H_0^2\Omega_A/a^2\gg \{a''/a, \; k^2\}$ is satisfied at some point of the propagation. As a matter of fact, since in the radiation-dominated epoch  $a''/a\propto 1/a$,  the new term can grow quickly as we go back in time and eventually become dominating in the early universe as can be seen in Fig. \ref{fig:evolution}

It is also worth noting that the vector field contribution is slightly larger in the plus equation and contains an additional $\sin^4 \theta$ anisotropic modulation compared to the cross equation, so any effect resulting from the vector field will be enhanced for this polarization. If the new term is subdominant, the mode will  behave the standard way as described in Section \ref{sec:GWs}. 

Another major point is the fact that this new contribution makes the GW phase velocity $c_T$ slightly different from the speed of light $c$. The relative difference between these two $\delta c=|c_T-c|/c$ has been strongly bounded by the gravitational-wave event GW170817 observed by the Advanced LIGO and Virgo detectors, and the gamma-ray burst (GRB) GRB 170817A  independently observed by Fermi  \cite{LIGOScientific:2017zic} and is limited to $\delta c\lsim\mathcal{O}(10^{-15})$ at the frequency operating range $f=10\text{ Hz}\sim 10\text{ kHz}$. The modified dispersion relation that appears in the equation of propagation, which can be seen either as an anomalous velocity for the tensor modes $c_T\neq 1$ or an effective mass of the graviton $m_g\neq 0$  \cite{Ezquiaga:2018btd}, yields a much smaller deviation, at about $\delta c \leq \mathcal{O}(10^{-44})$ for those frequencies and the aforementioned upper bound for $\Omega_A$. This difference in velocity between polarizations produces a Shapiro time delay as well as residuals for pulsars, although the effects are too small to be measured with current or near-future detectors. 

On the other hand, depending on the vector field oscillation frequency $\omega$, we can consider two regimes.  In the slow-oscillation regime
$\omega\ll {\cal H}$ both polarizations are affected, with the oscillating term being larger for the plus polarization. An analytical solution of the differential equation is not possible in this case.

If instead the oscillation of the field is fast enough ($\omega\gg  {\cal H}$), the elliptic function can be averaged leading to the following effective equations:
\begin{equation}\label{eq:hU1fast}
    h_{\lambda}''+2\mathcal{H}h_{\lambda}'+\left[k^2+\frac{4 H_0^2 \Omega_A \sin^4\theta}{a^2} \delta_{\lambda,+} \right]h_{\lambda}=0,
\end{equation}
where $\delta_{\lambda,+}$ is the Kronecker delta. So in the fast-oscillation regime, the behaviour of the cross polarization reduces to the standard propagation equation in $\Lambda$CDM  \eqref{eq:GWpropvacuum}, whereas the plus polarization is still affected. It is also worth noting that these equations are independent of $\omega$, so changing the frequency of the field has no effect as long as the fast oscillation condition is satisfied. This much simpler equation allows for an analytic solution when the vector field term dominates. Thus, in a radiation-dominated era, in which $a(\eta)=a_r\eta$, with $a_r\simeq H_0\sqrt{\Omega_R}$, the mode evolution reduces to
\begin{equation}
    h_+''+\frac{2}{\eta}h'_+ + \frac{4 H_0^2 \Omega_A \sin^4\theta}{a_r^2\eta^2}h_+ = 0,
\end{equation}
which has the following solution
\begin{equation}\label{eq:hplusanalytical}
    h_+(\eta) = C_1 \eta^{-\frac{1}{2}(1-\sqrt{1-4\xi})} + C_2 \eta^{-\frac{1}{2}(1+\sqrt{1-4\xi})},
\end{equation}
where $C_1,C_2$ are integration constants and
\begin{equation}
    \xi = \frac{4 H_0^2 \Omega_A \sin^4\theta}{a_r^2}\simeq 4\sin^4{\theta}\frac{\Omega_A}{\Omega_R}\ll 1.
\end{equation}

Thus, for fast-oscillating vector fields in the radiation-dominated era, super-Hubble GW plus modes undergo a slight damping with $h_+\propto a^{-\xi}$ in contrast with the constant behaviour of such modes
in $\Lambda$CDM. Oscillation would be possible as long as $\xi>1/4$, but the upper bound on $\Omega_A$ implies that $\xi<0.1$, which forbids it. A similar analysis can be done for a matter-dominated epoch, for which the long-term behaviour for the plus polarization is a constant value.

Fig. \ref{fig:evolution} shows the numerical evolution of three different modes alongside the radiation-dominated solution in the fast-oscillation regime. We can see the qualitative behaviour of the GW modes when each of the three terms in the propagation equation dominates:
\begin{enumerate}
    \item If the damping term $a''/a$ dominates, the mode is a purely super-Hubble mode so that it remains at a constant value.
    \item If the wavenumber term $k$ dominates, the mode is a purely sub-Hubble mode, which oscillates with its amplitude damped as $1/a$.
    \item If the vector field term dominates \emph{and} the fast-oscillation regime applies, the cross mode remains unaffected so it evolves according to whichever of the other two terms is dominating. The plus mode decays as $a^{-\xi}, \xi>0$ according to Eq. \eqref{eq:hplusanalytical}. A net GW polarization is thus generated during this stage.
\end{enumerate}

The change in the total power spectrum and the
generated non-zero linear polarization power spectrum can be described by means of $\mathcal{I}$ and $\mathcal{Q}$, as defined in Eq. \eqref{eq:reduced_stokes}. $\mathcal{U}=\mathcal{V}=0$ since the linear polarizations do not mix in the chosen basis ($R_{+\times}=R_{\times +}=0$). 

In our model, the power spectra are anisotropic, exhibiting a dependence on the polar angle $\theta$, so we shall perform a multipole decomposition of both non-zero reduced Stokes parameters as 
\begin{equation}\label{eq:multipolar_expansion}
    \mathcal{S}(k,\theta,\eta) = \sum_\ell\sqrt{\frac{2\ell+1}{2}}\mathcal{S}_\ell(k,\eta) P_\ell(\cos\theta),
\end{equation}
and the normalization is chosen so that $||\mathcal{S}||^2=\sum_\ell \mathcal{S}_\ell^2$, where the norm takes the standard form
\begin{equation}
    ||\mathcal{S}||^2=\int_{-1}^1\dd{\cos\theta}\mathcal{S}^2(\theta).
\end{equation}

Both $\mathcal{I}_\ell$ and $\mathcal{Q}_\ell$ vanish for odd $\ell$ since the GW propagation equation is invariant under the transformation $\theta\to\theta'=\pi-\theta$. On top of that, the modulation of the new term in the GW propagation equation is proportional to $\sin^2\theta$ and $\sin^4\theta$, so since the power spectra contain the square of the GW amplitudes, we expect a significant contribution coming from multipoles up to $\ell=8$.

Finally, let us define the \emph{degree of polarization} $\mathcal{D}$, which measures how polarized the GW background is. For that matter, we shall take into account that the Stokes parameters satisfy $I^2\geq Q^2 + U^2 + V^2$, with the equality holding when there is total polarization. Thus, an appropriate way to define the degree of polarization is
\begin{equation}\label{eq:deg_of_polarization}
    \mathcal{D}^2 = \frac{\mathcal{Q}^2+\mathcal{U}^2+\mathcal{V}^2}{||\mathcal{I}||^2},
\end{equation}
which can be decomposed into multipoles as well. In this case, it is given simply by
\begin{equation}
    \mathcal{D}_\ell = \frac{\mathcal{Q}_\ell}{\sqrt{\sum_l \mathcal{I}_\ell^2}},
\end{equation}
which gives a measure of how much linear polarization there is in each multipole. If the GW background is completely polarized, i.e. only one of the two polarizations occurs, then $\sum_\ell \mathcal{D}_\ell^2 = 1$, and if that configuration happens to be allocated only in a particular multipole $\ell = n$, then $\mathcal{D}_n = 1$, with all the other components vanishing.

\begin{figure}[t]
    \centering
    \includegraphics{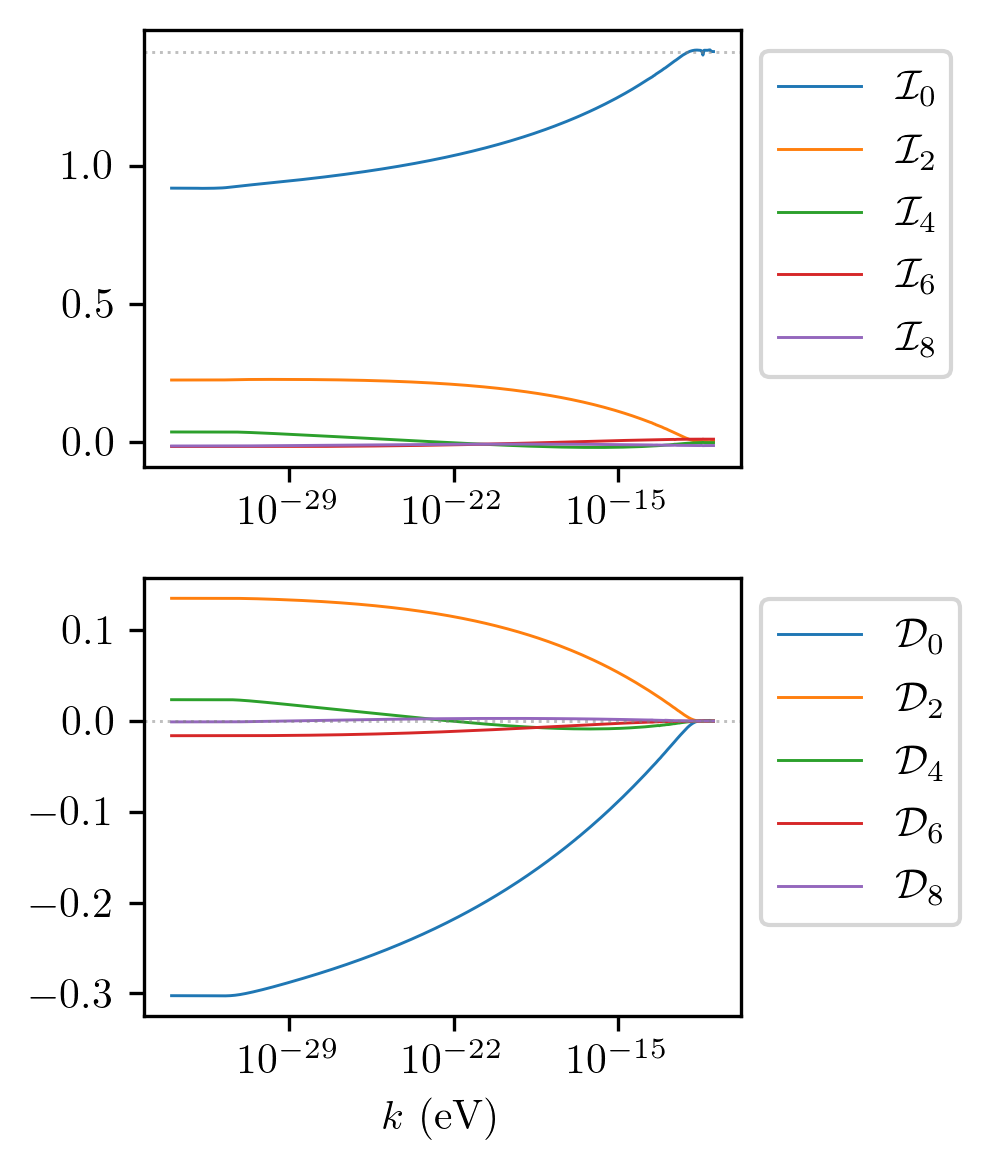}
    \caption{Multipole expansion up to $\ell=8$ of intensity and degree of polarization today, for $\Omega_A=\Omega_R/100$ and $\omega=2500$ Hz. The gray dotted line in the $\mathcal{I}_\ell$ plot is the value $\mathcal{I}_0=\sqrt{2}$, which corresponds to the case of no polarization and no deviation from standard cosmology $R_+=R_\times=1$.}
    \label{fig:stokes}
\end{figure}

Fig. \ref{fig:stokes} shows a particular example of this multipole decomposition for the different Stokes parameters evaluated today. We see the suppression of power in the monopole ${\cal I}_0$ at large scales with respect to $\Lambda$CDM similarly to that found for vector dark matter \cite{Miravet:2020kuj} and the generation of a quadrupole and hexadecapole contributions which are absent in $\Lambda$CDM, whereas higher multipoles are negligible. On the other hand, we also see  a large degree of polarization with a monopole  distribution and also non-negligible polarization with a quadrupolar  and hexadecapolar distribution patterns.  We can clearly see that a larger amount of net polarization happens for modes with smaller wavenumbers $k$, along with a diminution of the total power spectrum, both originated by the decay of the plus modes. Such an effect is due to the fact that modes with larger wavenumbers enter the Hubble horizon earlier in their evolution, so that they undergo the decay caused by the vector field for a smaller time. An extreme yet clear indicator of this is that modes with $k\gsim 10^{-11}$ eV have always been inside the Hubble horizon (for the chosen reheating scale $T_\text{RH}=10^{10}$ GeV), and as a result no difference from standard cosmology is observed for them, so that $R_+=R_\times=1$ (equivalently, $\mathcal{I}=1, \mathcal{Q}=0$). On the other side, we have  those modes with wavelengths larger than today's Hubble radius $k<H_0\simeq 10^{-33}$ eV so they have always been super-Hubble and have evolved in the same way 
irrespective of $k$, giving rise to the flat plateau observed in the 
low-$k$ region of the figure.

\begin{figure}[t]
    \centering
    \includegraphics{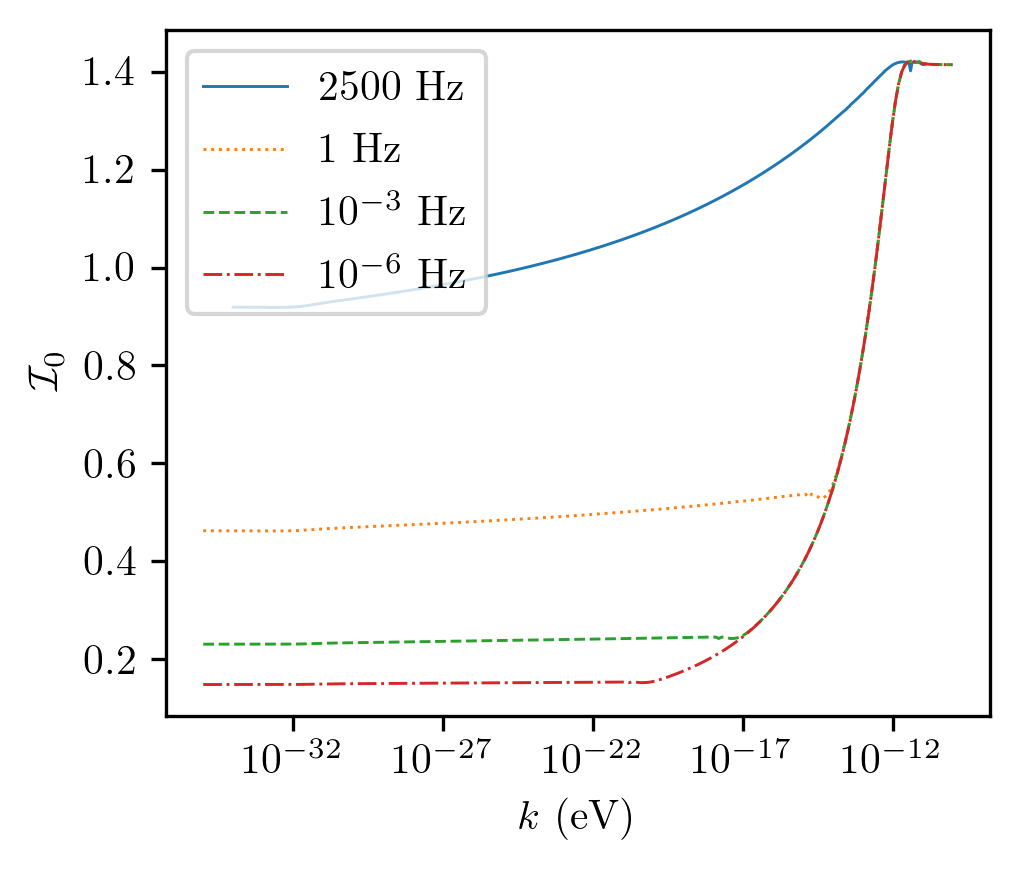}
    \caption{Monopole of the total power spectrum today $\mathcal{I}_0$ as a function of wavenumber, for $\Omega_A=\Omega_R/100$ and four different frequencies $\omega$ of the field. Smaller frequencies yield a larger suppression of the power spectrum, as a result of the field oscillating slowly for longer.}
    \label{fig:I0}
\end{figure}

If the frequency of the vector field is not large enough to be always in the fast  oscillation regime, the GW modes are affected by the slow oscillation of the vector field. Since $\omega\eta\ll 1$, the elliptic cosine in Eq. \eqref{eq:hU1} is approximately constant with value $\cn\simeq 1$. As a result, and as long as the modes are super-Hubble, they undergo a damping which is similar to that of the plus mode in the fast-oscillation regime, but steeper, as the vector field term is now slightly larger (even larger for the plus polarization). When the vector field enters the fast-oscillation regime, the cross polarization mode freezes and follows a standard propagation, whereas the plus mode keeps damping, with a less steep slope, until it enters the Hubble horizon.
This can be seen in Fig. \ref{fig:I0}, which shows that the monopole $\mathcal{I}_0$ is more suppressed for smaller frequencies, as both polarizations are damped for longer. All curves have the same behaviour  for wavenumbers $k\gg \omega$, i.e. for modes that enter the Hubble radius during the slow-oscillation phase
of the vector field. 
 This particular example aims just to illustrate the effect of a slow oscillation, since the anisotropy magnitude, about $\Omega_A/\Omega_R=1/100$,  is higher than the typical cosmological perturbations, and thus an accurate study would require a description in terms of a Bianchi I spacetime background.
 
\subsection{Circularly polarized vector field}\label{subsec:GWcirc}

\begin{figure}[t]
    \centering
    \includegraphics[width=60mm]{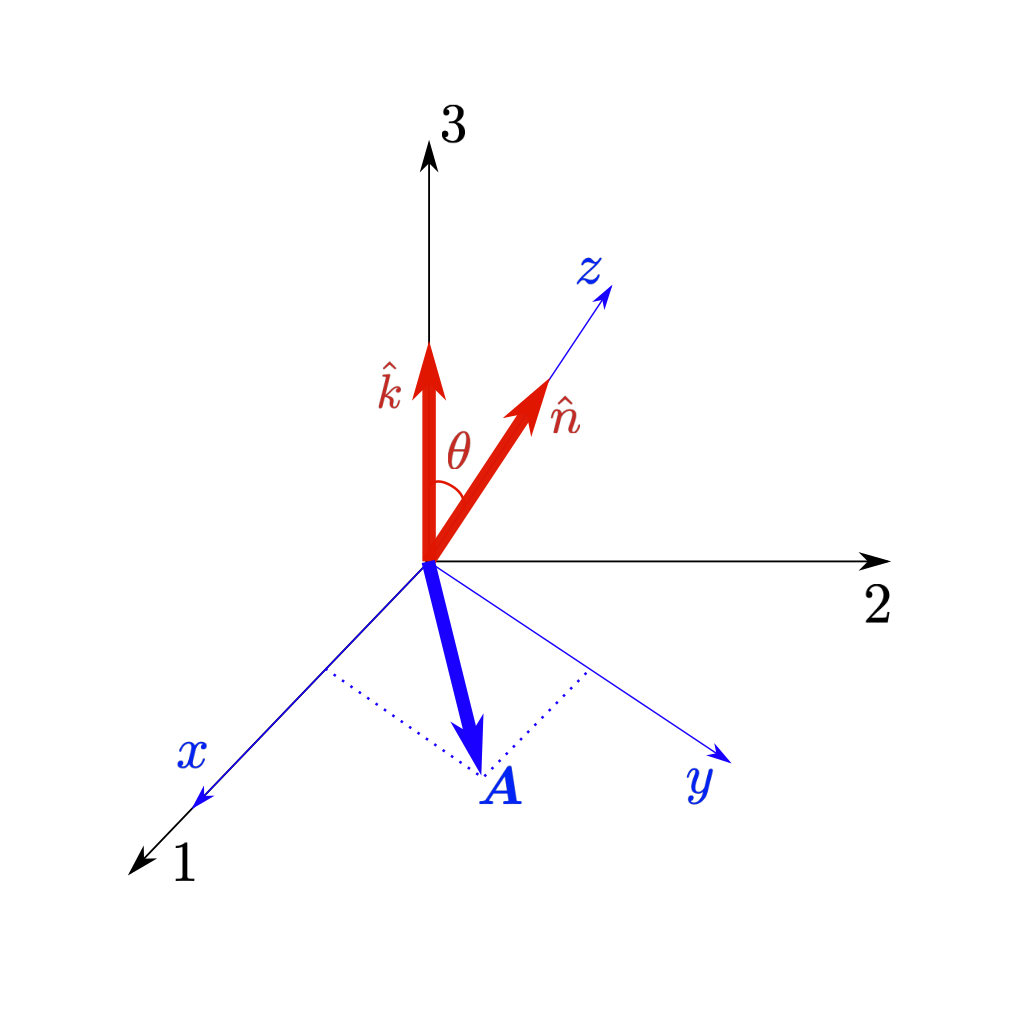}
    \caption{Coordinate systems arrangements for the circularly polarized vector field. The GW coordinate system is labelled $(1,2,3)$ so that the GW travels along the 3-direction, defined by the wavenumber $\vb{\hat{k}}$. The vector field coordinate system is labelled $(x,y,z)$, with the field rotating in the $xy$-plane, defined by its normal $\vb{\hat{n}}$, which forms an angle $\theta$ with the propagation direction. The GW plane is chosen so that directions $1$ and $x$ match.}
    \label{fig:circ_axis}
\end{figure}

We repeat now the same analysis with the circularly polarized vector field. We now label the GW propagation direction as ``3'' instead of ``z'' so as to avoid confusion with the $(x,y,z)$ system of coordiantes defined by the vector field. Thus, the GW propagates along the 3-axis and, thanks to axial symmetry, we choose the normal to the vector field rotation plane $\vb{\hat{n}}$ to be in the $23$-plane. In addition, we choose the vector field to be initially oriented towards the $1$-direction, so that $A_x$ as defined in Eq. \eqref{eq:A_circ} coincides with $A_1$ in this coordinate system. This layout is shown in Fig. \ref{fig:circ_axis}. The GW propagation equations are now given by
\begin{subequations}\label{eq:GWprop_circ}
\begin{equation}
    h_{+}''+2\mathcal{H}h_{+}' + k^2 h_{+} + \frac{2\Omega_A H_0^2}{a^2}[(F+B)h_{+} + M h_{\times}]=0,
\end{equation}
\begin{equation}
    h_{\times}''+2\mathcal{H}h_{\times}' + k^2 h_{\times} + \frac{2\Omega_A H_0^2}{a^2}[(F-B)h_{\times} + M h_{+}]=0,
\end{equation}
\end{subequations}
where
\begin{widetext}
\begin{subequations}
\begin{alignat}{1}
        F &= \frac{\lambda(A_1^2+A_2^2)(A_1^2+A_2^2+2\alpha^2)-2(A_1'{}^2+A_2'{}^2)}{\lambda\alpha^4}\nonumber \\
        &= \cos^4{\omega\eta} + \cos^4{\theta}\sin^4{\omega\eta} + 2\sin^2{\theta}\cos{2\omega\eta} + \frac{1}{2}\cos^2{\theta}\sin^2{2\omega\eta}, \\
        B &= \frac{\Re[(A_1+iA_2)^4]}{\alpha^4} = \cos^4{\omega\eta} + \cos^4{\theta}\sin^4{\omega\eta} - \frac{3}{2}\cos^2{\theta} \sin^2{2\omega\eta},\\
        M &= \frac{\Im[(A_1+iA_2)^4]}{\alpha^4} = 2\cos{\theta}\sin{2\omega\eta}(\cos^2{\omega\eta} - \cos^2{\theta}\sin^2{\omega\eta}),
\end{alignat}
\end{subequations}
\end{widetext}
and $\cos{\theta}=\vb{\hat{k}}\cdot \vb{\hat{n}}$. These equations exhibit some similarities with the linearly polarized case: The new terms are proportional to the abundance $\Omega_A$ and dominate in the early universe for modes with sufficiently small $k$ due to the $a^{-2}$ scaling, in which case they are expected to affect GWs in the early stages of their evolution. The equations are also different for each polarization, which produces a net polarization of the GW background. On top of that they are coupled, with each mode acting as a source of the other, which enhances the polarization generation, but this mixing is purely real, so parity is still preserved.

When $\omega\eta\gg 1$, a fast-oscillation regime applies, in which the oscillations of the vector field  can be averaged for the integration of the GW propagation, resulting in
\begin{alignat}{1}
    \langle F \rangle &= \frac{3}{8}\left(1+\cos^4{\theta}\right) + \frac{1}{4} \cos^2{\theta}, \\
    \langle B \rangle &= \frac{3}{8}\left(1+\cos^4{\theta}\right) - \frac{3}{4} \cos^2{\theta}, \\
    \langle M \rangle &= 0.
\end{alignat}

In this regime, the polarizations do not mix anymore, so if there is any mixing between both polarizations it needs to happen when the vector field is revolving slowly. As in the previous subsection, it is also possible to obtain the analytical solution for super-Hubble modes in the radiation-dominated era, which is given by \eqref{eq:hplusanalytical} for both polarizations with
\begin{equation}
    \xi_{\substack{+\\ \times}} = \frac{2\Omega_A H_0^2 \langle F \pm B\rangle}{a_r^2} \simeq 2\langle F \pm B\rangle\frac{\Omega_A}{\Omega_R}.
\end{equation}

Since the long-time behaviour of the modes is $h_\lambda\propto a^{-\xi_\lambda}$, and $\xi_+ > \xi_\times$, super-Hubble plus-polarized modes are more suppressed than cross-polarized ones, especially around $\theta = \pi/2$ where the difference is maximum.

\begin{figure}[t]
    \centering
    \includegraphics{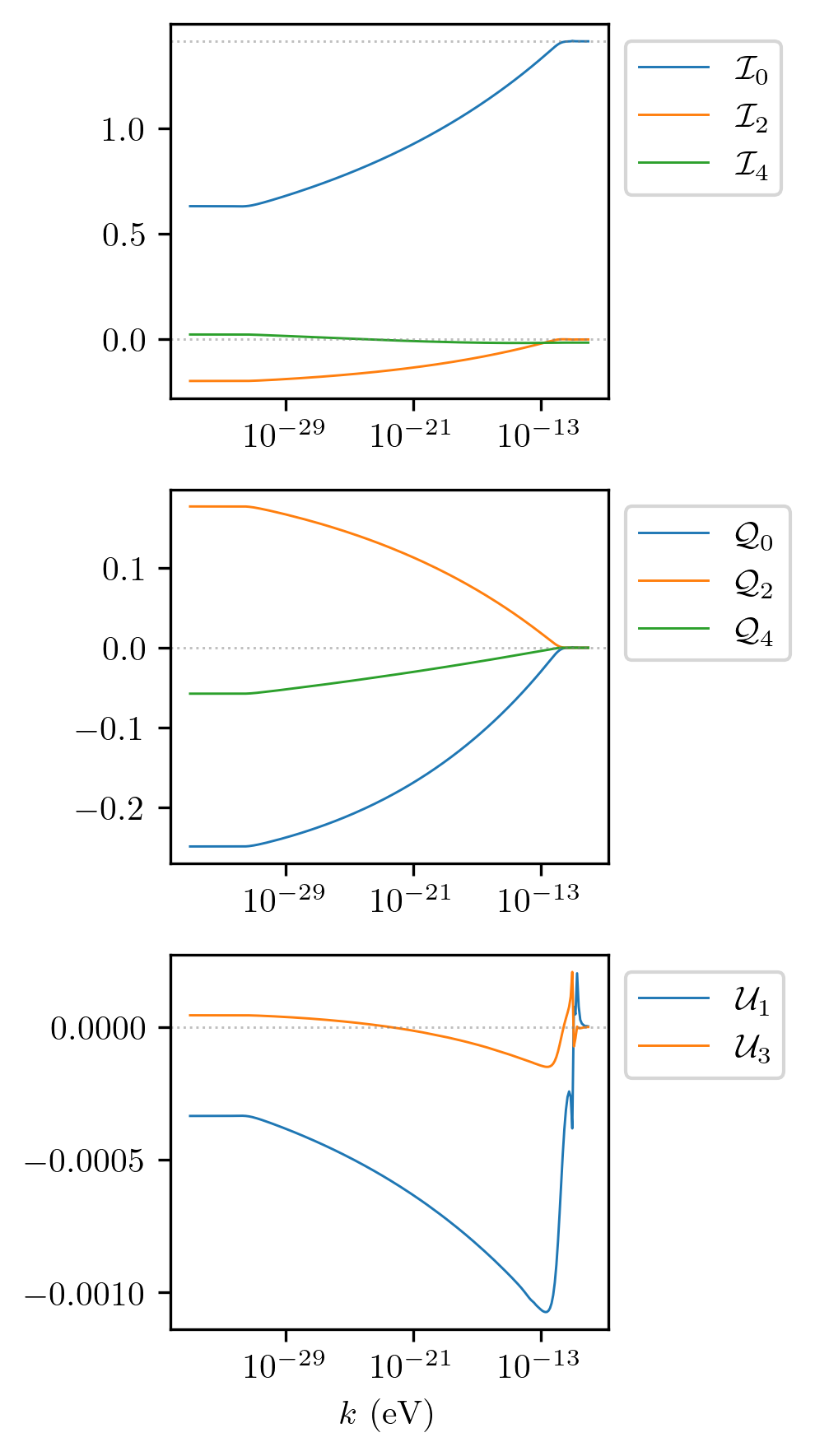}
    \caption{Multipole expansion of the reduced Stokes parameters today for the circularly polarized vector field, with $\Omega_A=\Omega_R/100$ and $\omega = 2500$ Hz. $\mathcal{I}$ and $\mathcal{Q}$ have even multipoles, whereas $\mathcal{U}$ has odd ones. The linear polarization is dominated by $\mathcal{Q}\gg\mathcal{U}$ except in the high-$k$ region, where the total spectrum remains unchanged $\mathcal{I}_0=\sqrt{2}$ but polarization still appears (see Fig. \ref{fig:stokes_circ_hk}).}
    \label{fig:stokes_circ}
\end{figure}

Let us look at the Stokes parameters now. Circular polarization still does not occur, therefore $\mathcal{V}=0$, but as opposed to the previous section, now both linear polarizations do mix, which implies $\mathcal{U}\neq 0$. Fig. \ref{fig:stokes_circ} shows all three non-zero Stokes parameters as a function of wavenumber for the same abundance and frequency as in Fig. \ref{fig:stokes}. Once again, $\mathcal{I}$ and $\mathcal{Q}$ receive contributions from even multipoles only. This is because the equations of propagation are different in a term proportional to $B$, which is even in $0\leq\theta\leq\pi$. However, the anisotropy created by the mixing of both polarizations, which is governed by $M$, is odd in this same interval, so $\mathcal{U}$ receives contribution from odd multipoles only.

\begin{figure}[t]
    \centering
    \includegraphics{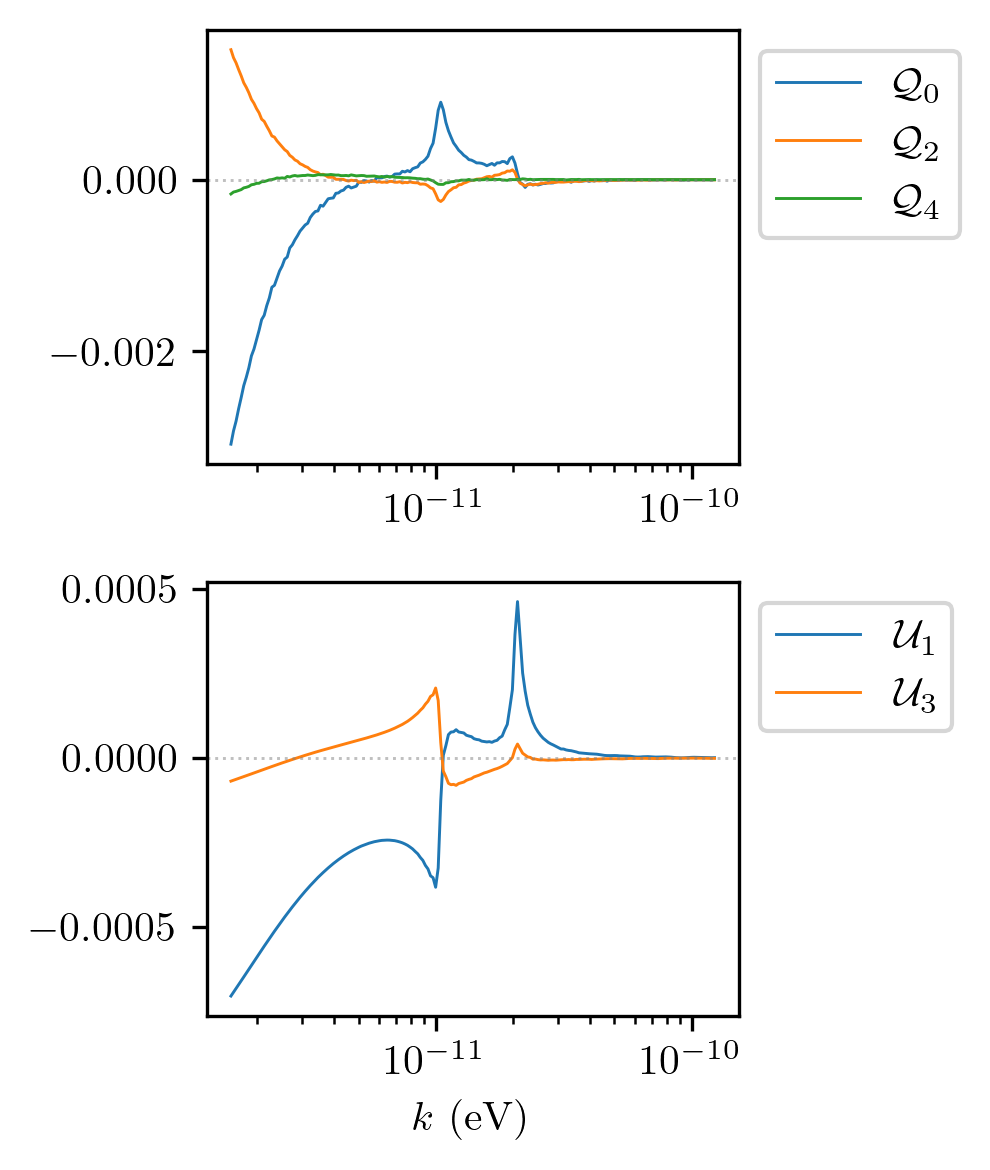}
    \caption{Stokes parameters $\mathcal{Q}$ and $\mathcal{U}$ today in the high-wavenumber region, smaller than the size of Hubble horizon at reheating, with $\mathcal{H}_\text{RH}\simeq 10^{-12}$ Hz. In this region, both Stokes parameters accountable for linear polarization are about the same size. The fundamental and first overtone of $k=\omega\simeq 10^{-11}$ eV are seen as resonances. For larger $k$, both parameters go to zero, as net linear polarization is no longer generated.}
    \label{fig:stokes_circ_hk}
\end{figure}

Despite this new source of anisotropy, the generation of net linear polarization is still dominated by the difference in the term that involves $F\pm B$ in Eq. \eqref{eq:GWprop_circ}, rather than the source term, since $\mathcal{Q}\gg \mathcal{U}$. The sourcing is only possible when the vector dark radiation is not oscillating rapidly, which cannot happen for a long period of time without breaking background isotropy. Similarly to the previous section, modes in larger scales are more suppressed and exhibit a larger degree of linear polarization. Comparatively, the total power spectrum suffers a greater diminution, since the circularly polarized vector field causes both GW polarizations to decay, instead of just one of them. As a result, the difference in propagation between polarizations is smaller, which causes the net polarization to be smaller as well.

The only region in $k$ where  $\mathcal{U}$, and therefore the mixing, become important corresponds to modes that initially (at the end of  reheating in this case) were sub-Hubble, i.e. $k\gsim \mathcal{H}_\text{RH}$. In that case, the polarizations do not undergo a different super-Hubble damping phase, so $\mathcal{Q}$ is suppressed, and at the same time $k$ is not large enough to completely dominate the evolution, which would make both polarizations evolve the same way. 

Fig. \ref{fig:stokes_circ_hk} shows a zoom of this high-wavenumber region. Besides the already mentioned suppression of $\mathcal{Q}$ at the scale of the Hubble horizon, two resonances at $k=\omega\simeq 10^{-11}$ eV and its first overtone are also apparent. For bigger wavenumbers, the $k$ term completely dominates the propagation and no sign of polarization is observed. Note that this discussion is valid for GWs of cosmological origin. Even though high-frequency GWs coming from astrophysical events (such as compact binaries or pulsars) lie around this region in wavelength, this effect and resonances would not be present, as they originate when the mode is well inside the Hubble horizon.

\begin{figure}[t]
    \centering
    \includegraphics{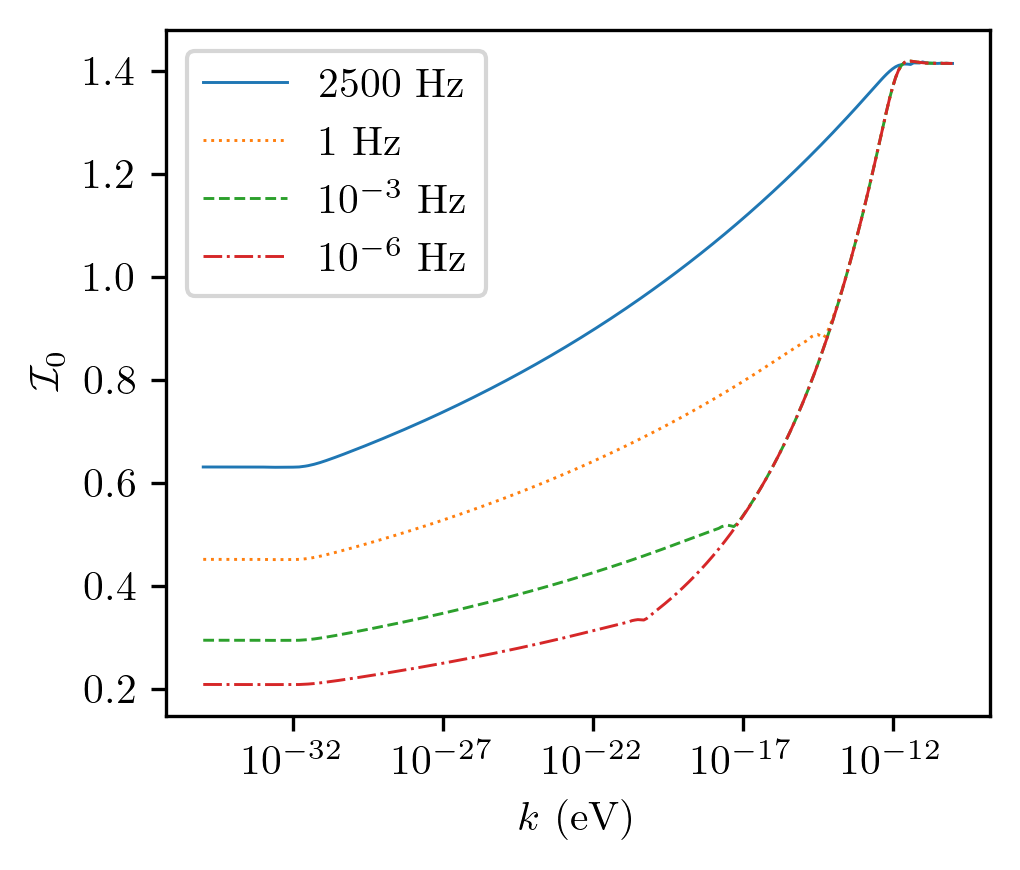}
    \caption{Monopole of the total power spectrum $\mathcal{I}_0$ today as a function of wavenumber, for $\Omega_A=\Omega_R/100$ and four different frequencies $\omega$ of the circularly polarized field. The behaviour is similar to that of the linearly polarized vector field, with smaller frequencies causing a larger suppression.}
    \label{fig:i0_circ}
\end{figure}

For the sake of completeness, we plot in Fig. \ref{fig:i0_circ} the monopole of the total power spectrum for different frequencies of the vector field, which exhibit the same behaviour as the ones discussed in the previous section.

\section{Polarized primordial background}\label{sec:polarized_bg}

Up to this point, we have studied the effect of the dark radiation vector field on the propagation of GWs originating from an unpolarized stochastic background. Although that is the standard assumption, there is also the possibility of having a primordial background with some degree of polarization. In order to explore such scenario, let us consider the extreme example of a background which is initially totally linearly polarized with $h_\times=0$. This in particular implies according to \eqref{eq:hLR} that $h_L=h_R$, i.e., for every right-handed tensor mode, there is a left-handed one with the same direction, amplitude and phase \cite{Gubitosi:2016yoq}.

In terms of analysing the propagation of the modes, the only difference is that there is no $\times$ polarization at origin nor $\times$ modes sourced by $+$ modes, which makes $R_{+ \times} = R_\times = 0$. For the linearly polarized vector field, since tensor polarizations do not mix, it just means that the total power spectrum is reduced by half and becomes completely polarized, as only one of the two polarizations contributes.

\begin{figure}[t]
    \centering
    \includegraphics{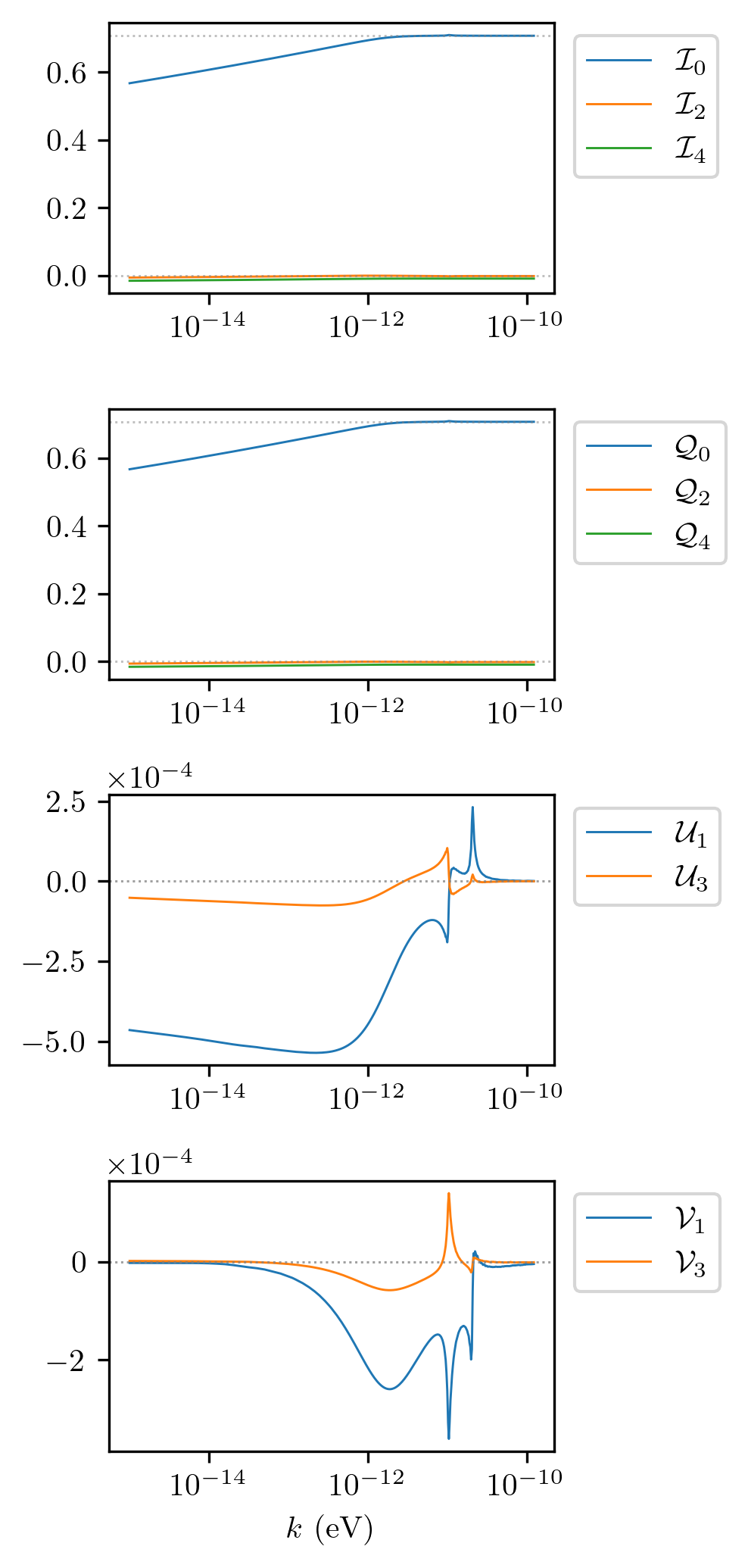}
    \caption{Multipole expansion of the reduced Stokes parameters today when the primordial background is initially totally linearly polarized. Such scenario allows for non-zero net circular polarization $\mathcal{V}$ generation in the high-wavenumber region, which also presents resonances at $k=\omega$ and its first overtone due to the polarization mixing.}
    \label{fig:stokes_v}
\end{figure}

For the circularly polarized vector field, GW polarizations are mixed, therefore the analysis is slightly different. The total power spectrum and the net linear polarization are still dominated by the fact that cross polarization is missing at origin. In terms of the reduced Stokes parameters \eqref{eq:reduced_stokes}, since $|R_+| \gg |R_{\times +}|$ as the sourced mode is considerably smaller than the source, $\mathcal{I} \simeq \mathcal{Q}$, which are roughly half as $\mathcal{I}$ with the unpolarized primordial background. 

The novelty is that this situation also opens up the possibility of creating net circular polarization. This requires the product $R_{+} R_{\times +}^*$ to have a non-vanishing imaginary part, i.e., that $h_+$ and $h_\times$ do not share the same phase. That happens just for modes which initially have $k\simeq {\cal H}$, which we have already discussed in the previous section. In this regime, the $h_\times$ mode, which is absent initially, needs to be produced before entering the Hubble horizon, which is what creates this phase delay between both polarizations and ultimately a circularly polarized realization. We plot in Fig. \ref{fig:stokes_v} all reduced Stokes parameters for this high-$k$ region. As stated before, $\mathcal{I}$ acquires half the values as in the unpolarized background, and $\mathcal{Q}$ is roughly equal to it. The other linear polarization $\mathcal{U}$, now much smaller than $\mathcal{Q}$, is also halved, as the contribution sourced by the $h_\times$ mode is not present, while still exhibiting two resonances, the fundamental one being at $k=\omega$. The new addition, the circular polarization parameter $\mathcal{V}$, is about the same order of magnitude as $\mathcal{U}$, and since it is also  generated by polarization mixing, it also has got the resonances and only odd-multipole contributions. As expected, the net circular polarization goes to zero as $k$ is so big that it completely dominates the evolution, and also as we go to very small $k$, where the difference in phase between polarizations disappears.

\section{GW propagation for a general vector potential}\label{sec:general_potential}

Lastly, let us consider an abelian vector field in a flat FLRW background with a general potential. The action for the vector field in this scenario is given by
\begin{equation}\label{eq:SVF}
    S_\text{VF} = \int \dd^4x \sqrt{-g} \left( -\frac{1}{4} F_{\mu\nu} F^{\mu\nu} + V(A^2)\right),
\end{equation}
with $V(A^2)$ the general self-interaction potential.

The stress-energy tensor for the vector field given by the following expression:
\begin{eqnarray}\label{eq:TVF}
    T^\alpha{}_\beta = \left(\frac{1}{4} F_{\mu\nu} F^{\mu\nu} + V(A^2)\right) \delta^\alpha{}_\beta\nonumber\\ 
    - F^{\alpha\mu} F_{\beta \mu} + 2V'(A^2) A^\alpha A_\beta, 
\end{eqnarray}
where the prime in $V'(A^2)$ denotes a derivative with respect to the argument of the potential. 

On the other hand, if we vary the action \eqref{eq:SVF} with respect to the vector field $A_\mu$, we get its equations of motion
\begin{equation}
    F^{\mu\nu}{}_{;\nu} - 2 A^\mu V'(A^2) = 0,
\end{equation}
which in components reads
\begin{subequations}
    \begin{equation}\label{eq:A0eom}
        A_0 V'(A^2) = 0,
    \end{equation}
    \begin{equation}\label{eq:Azeom}
        A_i'' + 2a^2 A_i V'(A^2) = 0.
    \end{equation}
\end{subequations}

Equation \eqref{eq:A0eom} requires $A_0 = 0$ in order to avoid a trivial solution in which the vector field would lie motionless at the minimum of its potential, which does not allow for an isotropic stress-energy tensor at background level. Therefore, we are left with just the spatial components $A_i$, whose evolution is determined by Eq. \eqref{eq:Azeom}.

Now we need to make an assumption about the field to allow for further analysis. A circularly polarized-like solution, i.e. the vector field revolving within a particular plane, cannot be readily studied for an arbitrary potential. The geometry of the motion will be more complicated in general, describing non-periodic or open trajectories that are unknown unless a particular potential is chosen,
so we restrict ourselves to a linearly polarized ansatz. Assuming a homogeneous vector field, we can write
\begin{equation}
    A_\mu(\eta) = (0, 0, 0, A_z(\eta)),
\end{equation}
and the solution for the only component left can be found after specifying the shape of the potential.

For tensor perturbations, we follow the same procedure as in Section \ref{sec:GWwithDR} to arrive to the modified propagation equations for GWs:
\begin{widetext}
\vspace{1cm}
\begin{subequations}
\begin{equation}
    h_{+}''+2\mathcal{H}h_{+}'+\left[k^2-\frac{8\pi G\sin^2\theta}{a^2}\left( \vb{A}'^2 - 2a^2 V'(A^2) \vb{A}^2 - 2V''(A^2) \vb{A}^4 \sin^2 \theta \right) \right]h_{+}=0,
\end{equation}
\begin{equation}
    h_{\cross}''+2\mathcal{H}h_{\cross}'+\left[k^2-\frac{8\pi G\sin^2\theta}{a^2}\left( \vb{A}'^2 - 2a^2 V'(A^2) \vb{A}^2  \right) \right]h_{\cross}=0.
\end{equation}
\end{subequations}
\vspace{1cm}
\end{widetext}

In these equations, $\theta$ is again the angle between the direction of propagation of the GW mode and the direction of the vector field, and $\vb{A}^2 = \delta^{ij}A_i A_j = A_z^2$, so that $\vb{A}^2 = -a^2 A^2$. The propagation for both polarizations is manifestly different provided that $V''(A^2)\neq 0$, which is true for any potential with the exception of constant or mass-like quadratic potentials and as in the quartic case, this implies the generation of a net linear polarization. Note that linear polarizations do not mix regardless of the potential, because since the vector field always points in the same direction, it is possible to align its transversal part with one of the two polarizations, and the net linear polarization occurs in that basis.

For a generic polarization of the vector field, we also expect  for
a general potential with $V''(A^2)\neq 0$ a similar phenomenology to that 
studied for a circularly polarized vector in the quartic case, with 
generation of $Q$ and $U$ polarization modes.  

\section{Conclusions}\label{sec:conclusions}

We have studied coherent vector fields with a quartic potential in an expanding universe, which play the role of a possible dark radiation component. We have analysed  their effects on GW propagation and, in particular, on the primordial GW background generated during inflation. We observe an overall suppression of the primordial GW background due to the effect of the vector field, which is relevant while the GW mode is super-Hubble, as a result of which the damping is larger for large-scale modes.
The suppression in GW intensity exhibits an anisotropic pattern whose angular power spectra contains only even multipoles.
The effect on GWs with astrophysical origin is negligible.

We have computed the Stokes parameters for the primordial GW background today, assuming it was initially Gaussian, isotropic and unpolarized. In the two studied cases (linearly and circularly polarized vector field), we find that a net linear polarization is generated, mainly due to the different damping of each of the two linear polarizations, which is observed on even multipoles only. A linearly polarized  vector field  causes a larger suppression of the polarization that is more aligned with the direction at which it points, but not a mixing between $+$ and $\times$ modes, thus generating only $Q$ polarization of the GW.  In the case of the rotating vector field, there is also a mixing between $+$ and $\times$
modes so that both $Q$ and $U$ polarizations with an anisotropic pattern are generated. Since $U$ is produced by the mixing of $+$ and $\cross$ polarizations, it contains odd multipoles only. This polarization generation is especially important at large scales.

We have also studied the case of a background which is initially totally polarized with pure linear polarization. We have found that for circularly polarized vector field, net GW circular polarization is produced for modes with wavelengths initially comparable to  the size of the Hubble horizon.

Next generation of CMB experiments, both ground-based such as BICEP Array \cite{Hui:2018cvg} or Simons Array \cite{Suzuki:2015zzg}, and satellite-based like LiteBIRD \cite{LiteBIRD:2022cnt}, with an improved  sensitivity for the measurement of tensor to scalar ratio $\sigma(r)<0.006$ (even smaller in the case of LiteBIRD $\sigma(r)<0.001$) will allow for the  detection of the primordial tensor modes generated in typical models of inflation with $r$ around $r=0.01$. This detection could take place through CMB B-mode observations for $\ell<200$ as larger multipoles are dominated by gravitational lensing. Therefore, with this sensitivity we expect that the effects resulting from vector dark radiation would be noticeable, mainly as an angular modulation of the tensor power spectrum, in that multipole range. Although it is possible to detect the polarization of a GW stochastic background with interferometers \cite{Kato:2015bye, Domcke:2019zls, Sato-Polito:2021efu}, the typical frequency range covered by this type of detectors is far away from those in which linear GW polarization is generated. However, 
the resonances observed in the $Q$ and $U$ parameters in Fig. \ref{fig:stokes_circ_hk} appear 
in the detectable frequency range for certain values of the vector field
oscillation frequency, although for typical primordial power spectra, 
the corresponding amplitude would be negligibly small. 

There are other effects which could also affect the propagation of GWs and that, in some cases, could be degenerate with the presence of vector dark radiation. In particular, neutrino free streaming explored first in \cite{Weinberg:2003ur} produces an anisotropic stress which induces a  damping of GWs. Decoupled neutrinos induce a suppression in the amplitude of GWs ranging from $5\%$ to $20\%$ for the modes that enter the Hubble horizon well after the neutrinos decouple from the photons, which corresponds to scales $k\ll 10^{-25}$ eV. Although the neutrino-induced damping is degenerate in this wavenumber region with the monopole suppression produced by vector dark radiation, it does not feature an anisotropic suppression  nor a polarization generation of the gravitational wave background and could be observationally disentangled.

Although in this work we have limited ourselves to abelian vector fields, conformal vector models based on non-abelian fields can also be considered
as dark radiation. These models exhibit  a richer phenomenology since, as shown in \cite{Caldwell:2018feo,Jimenez:2019lrk}, gravitational wave oscillations between different tensor modes would be possible. These models will be explored elsewhere. 

\acknowledgements{We would like to thank Jose Beltr\'an for helpful discussions. This work has been supported by the MINECO (Spain) project PID2019-107394GB-I00 (AEI/FEDER, UE). A.D.M. acknowledges financial support by the MICIU (Spain) through a Formación de Profesorado Universitario (FPU) fellowship FPU18/04599.}

\bibliography{paper-gwdr}

	\end{document}